\documentclass[11pt]{revtex4-1}

\usepackage{graphicx}
\usepackage{dcolumn}
\usepackage{color}
\usepackage{bm}
\usepackage{amsmath}
\usepackage{amsfonts}
\usepackage{amssymb}

\input epsf

\begin{document}

\title{Dynamic wormholes with particle creation mechanism}

\author{Supriya Pan\footnote{span@research.jdvu.ac.in}}
\affiliation{Department of Mathematics, Jadavpur University, Kolkata-700 032, India.}

\author{Subenoy Chakraborty\footnote{schakraborty@math.jdvu.ac.in}}

\affiliation{Department of Mathematics, Jadavpur University, Kolkata-700 032, India.}{}

\begin{abstract}
The present work deals with a spherically symmetric space--time which is asymptotically (at spatial infinity) FRW space--time and represents wormhole configuration: The matter component is divided into two parts---(a) dissipative but homogeneous and isotropic fluid, and (b) an inhomogeneous and anisotropic barotropic fluid. Evolving wormhole solutions are obtained when isotropic fluid is phantom in nature and there is a big rip singularity at the end. Here the dissipative phenomena is due to the particle creation mechanism in non-equilibrium thermodynamics. Using the process to be adiabatic, the dissipative pressure is expressed linearly to the particle creation rate. For two choices of the particle creation rate as a function of the Hubble parameter, the equation of state parameter of the isotropic fluid is constrained to be in the phantom domain, except in one choice, it is possible to have wormhole configuration with normal isotropic fluid. \\

Keywords: Wormholes, Particle creation, Isotropic, Equation of state.\\

Pacs No:  04.20.Jb, 04.70.Dy, 11.10.Kk

\end{abstract}
\maketitle
\section{Introduction}
A wormhole is an imaginary intuitive concept in general relativity. It acts like a bridge or tunnel to connect two or more asymptotic regions. However, this hypothetical object has become one of the most popular and intensively studied research area in general relativity. The studies so far in this topic can be divided into two classes: static wormholes and dynamic wormholes. Although, there are static wormhole solutions \cite{Ellis1, Bronnikov1} since 1973, but, the work by Morris and Thorne \cite{Morris1} has the key role in studying the static wormholes. Usually, the static wormhole space--time is sustained by a single fluid component which requires the violation of the null energy condition (NEC) \cite{Visser1, Hochberg1, Visser2, Ida1}. However, in asymptotically flat space--time, this violation of NEC is a consequence of the topological censorship \cite{Friedman1}. Most of the studies in wormholes are related to traversable wormholes which have no horizons, and as a result, there is two way passage through them. Although, the speed of light is not locally surpasses \cite{Visser3}, but due to global space--time topology \cite{Visser1, Visser3, Lobo1}, it is possible to have superluminal travel through these wormholes, and as a result, there is the idea of time machines \cite{Morris2, Visser4}\\

Further, it should be noted that, it is possible to construct wormhole space--times with an arbitrarily small violation of the averaged NEC \cite{Visser3}. So, it is speculated that, the wormhole configuration could be realized merely by quantum effects violating the energy conditions.\\

In general, wormhole geometries are not constructed by solving Einstein field equations, rather one first fixes the form of the space--time metric (i.e., redshift and shape functions) and then matter part is evaluated by computing the field equations. Due to Bianchi identities, the matter part so obtained automatically obey the local conservation equations and violates the NEC \cite{Morris1, Visser1, Visser2, Morris2, Dadhich1}. It should be noted that, in modified gravity theories, there are examples of traversable wormhole solutions without any violation of energy conditions, for example in Einstein--Gauss--Bonnet gravity \cite{Maeda1} and in higher dimensional Lovelock theories \cite{Matulich1}. Also, in this context, there are well known non-static Lorentzian wormholes in Einstein gravity where matter component may obey weak energy condition (WEC) but life time may be arbitrarily small, or, large intervals of time \cite{Kar1}.\\

On the other hand, dynamical wormholes (i.e., evolving relativistic wormholes \cite{Cataldo1}) are not as popular as static wormholes and also not well understood. The pioneering work related to dynamical wormholes was done independently by Hochberg and Visser \cite{Hochberg2} and Hayward \cite{Hayward1}. They independently choose quasi local definition of wormhole throat in a dynamical space--time. Essentially, wormhole throat is a trapping horizon \cite{Hayward2} of different kind but matter in both of them violates the NEC. However, Maeda, Harada, and Carr \cite{Maeda1} have shown another class of dynamical wormholes (cosmological wormholes) which are asymptotically Friedmann universe with a big bang singularity at the beginning. This class of wormholes do not need matter which violates NEC rather the dominant energy condition is satisfied everywhere. The basic difference between these two class of dynamical wormholes is purely from geometrical aspect. In the former case, wormhole throat is a 2D surface of non-vanishing minimal area on a null hypersurface, while, in the later class of wormholes, due to initial singularity, there is no past null infinity \cite{Maeda2}. As a result, the wormhole throat is defined only on a space--like hypersurface. Hence, there is no trapping horizon rather the space--times are trapped everywhere \cite{Maeda2}. In recent years, there are works with dynamic wormhole space--time filled with two fluids \cite{Cataldo2, Cataldo3}. Such a matter system is very much relevant in present day cosmology where such two fluid models are widely used to describe the observed accelerated expansion of the universe \cite{Cataldo4, Pan1}.\\

In the present work, we make an attempt to find dynamical wormhole solutions for two fluid system, where one is a dissipative homogeneous fluid, and the other fluid component is anisotropic and inhomogeneous in nature. We assume that the dissipation arises due to particle creation mechanism in non-equilibrium thermodynamics which for simplicity is assumed to be isentropic in nature. As a result, the dissipative pressure is linearly related to the particle creation rate \cite{Pan2, Chakraborty1}. The paper is organized as follows: a review of earlier works on wormhole configuration with two non-interacting fluids has been done in section II. In section III, particle creation mechanism in the non-equilibrium thermodynamic prescription has been presented. Section IV shows possible wormhole solutions for different choices of the particle creation parameter. The paper ends with a brief overview in section V.\\

\section{Basic Equations: A review of earlier works}

The metric ansatz for the dynamic wormhole space--time is given by

\begin{equation}
ds^2= -e^{2 \Phi(r, t)} dt^2+ a^2 (t) \left[\frac{dr^2}{1-\frac{b(r)}{r}-K r^2}+r^2 d\Omega_2 ^2\right],
\end{equation}

where $\Phi (r, t)$ is the redshift function; $a(t)$ is the scale factor of the wormhole universe; $b(r)$ is the usual shape function for the wormhole; $d\Omega_2 ^2= d \theta^2+ \sin^2 \theta d \phi^2$, and $K$ takes values 0, $\pm 1$. In particular, if $\Phi (r, t)\rightarrow \Phi (r)$ and $a(t)$ $\rightarrow$ $a_0$, a constant, then the above metric ansatz describes a static wormhole universe, while metric (1) describes a FRW model, if $\Phi (r, t)= 0= b(r)$.\\

Suppose the matter distribution of the wormhole universe is described by two non-interacting fluid components (termed as Fluid I and Fluid II) together with a cosmological constant $\Lambda$. Fluid I is homogeneous and isotropic, but dissipative in nature having energy-momentum tensor:

\begin{equation}
T_{\mu\nu}^I= (\rho+ p+ \Pi) u_\mu u_\nu+ (p+ \Pi) g_{\mu \nu},
\end{equation}

where $\rho= \rho (t)$ is the energy density, $p= p (t)$ and $\Pi$ are the isotropic pressure and the pressure due to dissipation respectively and $u_\mu$ is the four velocity of the fluid. On the other hand, Fluid II is both inhomogeneous and anaisotropic in nature with the energy-momentum tensor:

\begin{equation}
T_{\mu\nu}^{II}= (\rho_{in}+ \rho_t) v_\mu v_\nu+ p_t g_{\mu \nu}+ (p_r- p_t) \chi_\mu \chi_\nu,
\end{equation}

where, $\rho_{in}= \rho_{in} (t, r)$ is the energy density of the inhomogeneous fluid component, the anisotropic pressure is characterized by radial and transversal components by $p_r= p_r (t, r)$, and $p_t= p_t (t, r)$ respectively ($p_r= p_t$ implies that Fluid II is isotropic but inhomogeneous in nature), and $v_\mu$ and $\chi_\mu$ are respectively unit time--like and space--like vectors, i.e., $v_\mu v^\mu= -\chi_\mu \chi^\mu= -1$, $\chi^\mu v_\mu= 0$. Thus, the explicit form of the Einstein's field equations:

\begin{equation}
G_{\mu \nu}= -\kappa \left(T_{\mu\nu}^I+ T_{\mu\nu}^{II}\right)- \Lambda g_{\mu \nu},
\end{equation}

are \cite{Cataldo2}

\begin{equation}
3 e^{-2 \Phi(r, t)} H^2+ \frac{b^\prime}{a^2 r^2}+ \frac{3 K}{a^2}= \kappa \rho_{in}+ \kappa \rho +\Lambda,
\end{equation}

\begin{eqnarray}
-e^{-2 \Phi(r, t)} \left(\frac{2 \ddot{a}}{a}+ H^2\right)+ \frac{K}{a^2}- \frac{b}{a^2 r^3}+ 2e^{-2 \Phi(r, t)} H \frac{\partial \Phi}{\partial t}
&+&
\nonumber
\\
 \frac{2}{r^2 a^2} (r-b) \frac{\partial \Phi}{\partial r}= \kappa p_r+ \kappa (p+ \Pi)- \Lambda,
\end{eqnarray}

\begin{eqnarray}
e^ {-\Phi(r, t)} \left(\frac{2 \ddot{a}}{a} + H^2\right)+ \frac{K}{a^2}+ \frac{b-r b^\prime}{2 a^2 r^3}+ 2 e^{-\Phi(r, t)} H \frac{\partial \Phi}{\partial t}+ \left(\frac{2r-b-rb^\prime}{2a^2 r^2}\right) \frac{\partial \Phi}{\partial r}
&+&
\nonumber \\
\frac{r-b}{a^2 r} \left[\left(\frac{\partial \Phi}{\partial r}\right)^2+ \frac{\partial^2 \Phi}{\partial^2 r}\right]
=\kappa p_t +\kappa (p+ \Pi)- \Lambda,
\end{eqnarray}

and

\begin{equation}
2 \dot{a} e^ {-\Phi(r, t)} \left(\sqrt{\frac{r-b(r)}{r}}\right) \frac{\partial \Phi (r, t)}{\partial r}= 0.
\end{equation}

where $\kappa= 8 \pi G$, $u^\alpha= (e^{-\Phi}, 0, 0, 0)$ is the time--like vector denoting the four velocity of both the fluids, $H$ $=$ $\dot{a}$/$a$ is the Hubble parameter, and an `overdot', or, a `prime' denotes the differentiation with respect to the cosmic time `$t$', or, the radial co-ordinate $r$ respectively. From the field Eq. (8), we see that two classes of solutions are possible, namely,

$$(I)~~\dot{a}=0~~~~~~(\mbox{static});~~~~~~~ (II)~~~~\frac{\partial \Phi}{\partial r}= 0~~~~~~~(\mbox{non-static}).$$

So in the present work, we shall consider only the second choice for dynamic wormhole solutions. As a result, (without any loss of generality) we can choose $\Phi (r, t)= 0$ (a rescaling of the time co--ordinate). Thus the wormhole metric (1) now simplifies to

\begin{equation}
ds^2= -dt^2+ a^2(t) \left[\frac{dr^2}{1-\frac{b(r)}{r}-K r^2}+r^2 d\Omega_2 ^2 \right].
\end{equation}

Now, due to non-interacting nature of the two fluids, both of them satisfy the conservation equations separately as

\begin{equation}
\frac{\partial \rho}{\partial t}+ 3H (\rho+ p+ \Pi)= 0,
\end{equation}

\begin{equation}
\frac{\partial \rho_{in}}{\partial t}+ H (3 \rho_{in}+ p_r+ 2 p_t)= 0,
\end{equation}

and

\begin{equation}
\frac{\partial p_r}{\partial r}= \frac{2}{r} (p_t- p_r).
\end{equation}

Here, Eq. (10) is the conservation equation for the homogeneous but dissipative fluid (i.e., Fluid I). Eq. (11) and Eq. (12) are the conservation equations for the other fluid component (i.e., Fluid II). Note that Eq. (12) is nothing but the relativistic Euler equation.

One may notice that the anisotropic nature (i.e., $p_t \neq p_r$) of the inhomogeneous fluid (i.e., Fluid II) is essential, otherwise, the pressure components become homogeneous as seen from Eq. (12), and then the conservation Eq. (11) demands that the density is also homogeneous. So, essentially, we have two non-interacting homogeneous fluid components leading a physically uninteresting situation. Hence, Fluid II must be anisotropic (i.e., $p_t \neq p_r$), and thus inhomogeneous.\\

We now write down the simplified form of the field Eqs. (5)--(7) for dynamical wormhole solution as \cite{Cataldo2, Cataldo3}

\begin{equation}
3 H^2+ \frac{3K}{a^2}+ \frac{b^\prime}{a^2 r^2}= \kappa \rho+ \kappa \rho_{in}+ \Lambda,
\end{equation}

\begin{equation}
-\left(2 \dot{H}+ 3 H^2+ \frac{K}{a^2}\right)-\frac{b}{a^2 r^3}= \kappa (p+ \Pi)+ \kappa p_r- \Lambda,
\end{equation}

and

\begin{equation}
-\left(2 \dot{H}+ 3 H^2+ \frac{K}{a^2}\right)+ \frac{b-r b^\prime}{2 a^2 r^3}= \kappa (p+ \Pi)+ \kappa p_t- \Lambda.
\end{equation}

Now, in order to solve the above non-linear field equations, we assume for simplicity that the radial and the transversal pressure components of Fluid II satisfy barotropic equation of state \cite{Cataldo3}:

\begin{equation}
p_r (t, r)= \omega_r \rho_{in},~~~~\mbox{and},~~~~~p_t (t, r)= \omega_t \rho_{in},
\end{equation}

where the constants $\omega_r$ and $\omega_t$ stand for the equation of state parameters. Now, inserting the relations in Eq. (16) to the conservation Eqs. (11) and (12), one immediately gets

\begin{equation}
\rho_{in} (t, r)= \rho_0 \frac{r^{\frac{2 (\omega_t- \omega_r)}{\omega_r}}}{a^{3+ \omega_r+ 2 \omega_t}},~~(\rho_0= \mbox{constant of integration}).
\end{equation}

Now, comparing the field Eqs. (14) and (15) and using Eq. (17), the shape function $b(r)$ can be obtained as

\begin{equation}
b(r)= K_0 r^3-\kappa \rho_0 \omega_r r^{-\frac{1}{\omega_r}},
\end{equation}

provided, we assume that the equation of state parameters are not independent, rather they are related by the following relation

\begin{equation}
\omega_r+ 2 \omega_t+ 1= 0.
\end{equation}

In the above solution for the shape function $b (r)$, the integration constant $K_0$ behaves as the curvature constant ($K$) in the metric shown in Eq. (1) or in Eq. (9). So, without any loss of generality, this integration constant $K_0$ may be absorbed by rescaling the radial co-ordinate `$r$' as follows:\\

$$K+ K_0= 1,~~~~~~~~~~~~~\mbox{when},~~K+K_0> 0.$$  $$K+ K_0= -1,~~~~~~~~~~~~~\mbox{when},~~K+K_0< 0.$$
$$K+ K_0= 0,~~~~~\mbox{for},~~(K= 1, K_0= -1; K= -1, K_0= 1; K=0, K_0= 0).$$

The gravitational configuration is described by the metric ansatz

\begin{equation}
ds^2= -dt^2+ a^2 (t)\left[\frac{dr^2}{1-K r^2-(\frac{r}{r_0})^{-\frac{1+ \omega_r}{\omega_r}}}+ r^2 d\Omega_2 ^2\right],
\end{equation}

there are two non-interacting fluid system in which the anisotropic and inhomogeneous matter component has energy density

\begin{equation}
\rho_{in} (t, r)= \frac{\rho_0 r^{-3-\frac{1}{\omega_r}}}{a^2}= -\frac{(\frac{r}{r_0})^{-3-\frac{1}{\omega_r}}}{\kappa a^2 \omega_r r_0 ^2},
\end{equation}

and the thermodynamic pressures along radial and transverse directions are

\begin{equation}
p_r (t, r)= \omega_r \rho_{in},~~~\mbox{and},~~~~p_t (t, r)= -\frac{1}{2} (1+ \omega_r) \rho_{in},
\end{equation}

while the homogeneous and isotropic part is described by the Friedmann equations

\begin{equation}
3\left(H^2+ \frac{K}{a^2}\right)= \kappa \rho + \Lambda,
\end{equation}

and

\begin{equation}
-\left(2 \dot{H}+ 3 H^2+ \frac{K}{a^2}\right)= \kappa (p+ \Pi)- \Lambda,
\end{equation}

and the fluid components are related by the conservation Eq. (10). Thus, for the present dynamic wormhole universe, the rate of expansion of these evolving wormholes is fully characterized by the homogeneous and isotropic, but dissipative matter component. In the following Section, we shall take an attempt to find the wormhole solutions, when dissipative phenomena is caused by the non-equilibrium thermodynamics due to particle creation mechanism.

\section{Non-equilibrium thermodynamics due to Particle creation}

As the number of particles is not conserved in non-equilibrium thermodynamic prescription, so the conservation equation for particle number takes the form \cite{Harko1}

\begin{equation}
\dot{n}+ 3 \theta n= n \Gamma,
\end{equation}

where $n= N/V$, is the particle number density; $N$ is the total number of particles in a co-moving volume $V$; $N^\mu= n u^\mu$ is the particle flow vector; $\theta= u^\mu _{;\mu}$ stands for fluid expansion; $\Gamma$ represents the particle creation rate, and notationally, $\dot{n}= n_{; \mu} u^\mu$. The sign of $\Gamma$ indicates creation ($\Gamma> 0$), or, annihilation ($\Gamma< 0$) of particles and $\Gamma$ represents some dissipative effect to the Cosmic fluid, so that, non-equilibrium thermodynamics comes into picture.\\

Using Clausius relation, the Gibb's equation takes the form \cite{Harko1}

\begin{equation}
T ds= d \left(\frac{\rho}{n}\right)+p d \left(\frac{1}{n}\right),
\end{equation}

where `$s$' is the entropy per particle and $T$ is the fluid temperature. Now, using the conservation relations (10) and (25), the entropy variation can be expressed as \cite{Zimdahl1}

\begin{equation}
n T \dot{s}= -\Pi \theta- \Gamma \left(\rho+ p\right).
\end{equation}

If for simplicity, we assume the thermal process to be `adiabatic' (or, `isentropic', i.e., $\dot{s}= 0$), then from the above equation we have

\begin{equation}
\Pi= -\frac{\Gamma}{\theta} \left(\rho+ p\right).
\end{equation}

Hence the dissipative pressure is completely characterized by the particle creation rate for the above isentropic thermodynamical system. Alternatively, the fluid may be considered as perfect fluid with barotropic equation of state: $p= (\gamma- 1) \rho$, and, dissipative phenomena comes into picture through particle creation. Note that, although the entropy per particle is constant but still there is entropy generation due to particle creation, i.e., enlargement of the phase space due to expansion of the universe. So, non-equilibrium configuration is not the conventional one due to the effective bulk pressure, rather a state with equilibrium properties as well (but not the equilibrium era with $\Gamma= 0$). Now, we eliminate $\rho$, $p$ and $\Pi$ from the Einstein field Eqs. (23), (24), and the isentropic Eq. (28), and then using barotropic equation of state parameter $\gamma= 1+ p/\rho$, we obtain \cite{Zimdahl1}

\begin{equation}
\frac{\Gamma}{3H}= 1+ \frac{2}{3 \gamma} \left(\frac{\dot{H}-\frac{K}{a^2}}{H^2+ \frac{K}{a^2}-\frac{\Lambda}{3}}\right).
\end{equation}

So, Eq. (29) helps us to conclude that for adiabatic thermodynamical system, the particle creation rate is related to the evolution of the universe.

\section{Evolving Wormhole Solutions}

In this section, we shall determine evolving wormhole solutions choosing the particle creation rate ($\Gamma$) as a function of the Hubble parameter ($H$). For simplicity, we take the flat model (i.e., $K= 0$) of the Universe.\\

From the field equations (23) and (24), the acceleration equation takes the form

\begin{equation}
\frac{\ddot{a}}{a}= -\frac{\kappa}{6} \left[\rho+ 3\left(p+ \Pi\right) \right]+ \frac{\Lambda}{3}.
\end{equation}

Hence, for expansion with constant velocity, we must have

\begin{equation}
\Lambda= \frac{\kappa}{2} \left[\rho+ 3 (p+ \Pi)\right],
\end{equation}

or, using Eq. (28) for `isentropic' condition, the particle creation rate for uniform velocity is restricted by

\begin{equation}
\Gamma= \frac{3 H}{\gamma} \left[(3 \gamma- 2)- \frac{2 \Lambda}{\kappa \rho}\right].
\end{equation}

So, for $\Lambda=0$ (or, $\Lambda \propto H^2$) we have $\Gamma \propto \rho^{\frac{1}{2}}$ (or, $\Gamma \propto H$). Also, for $\Lambda= 0$, positivity of $\Gamma$ restricts $\gamma$ to: $\gamma> 2/3$ or $\gamma< 0$; i.e., the homogeneous and isotropic fluid must be a normal fluid (satisfying strong energy condition) or is in phantom domain (i.e., violating weak energy condition) while there will be particle annihilation if the homogeneous fluid is in the quintessence era.\\

To evaluate the solutions, we start with the evolution Eq. (24) for $K= 0$, $\Lambda= 0$ and $\kappa= 1$, and using Eq. (28) for $\Pi$, we have

\begin{equation}
2 \dot{H}+ 3 \gamma H^2- \gamma H \Gamma= 0,
\end{equation}

which can be integrated to give

\begin{equation}
H= \frac{\exp \left({\frac{\gamma}{2} \int \Gamma dt}\right)}{H_0+ \frac{3 \gamma}{2} \int \exp \left({\frac{\gamma}{2} \int \Gamma dt}\right)dt}.
\end{equation}

Here $H_0$ is the constant of integration. The scale factor evolves as

\begin{equation}
a= a_0 \left[H_0+ \frac{3 \gamma}{2} \int \exp \left({\frac{\gamma}{2} \int \Gamma dt}\right)dt\right]^{\frac{2}{3 \gamma}};~~~~~~(a_0= \mbox{constant of integration}).
\end{equation}

We shall now determine the explicit solutions for the following choices of $\Gamma$.\\

$~~~~~~~~~~~~~~~~~~~~~~~~~~~~~~~~~~~~~~~~~~~~~~~~~~${{\bf Case I:} $\Gamma= \Gamma_0$,~a~constant}

The above solutions can explicitly be written as

\begin{eqnarray}
H&= &\exp \left(\frac{\gamma \Gamma_0 t}{2}\right) \left[H_0+ \frac{3}{\Gamma_0} \exp \left(\frac{\gamma \Gamma_0 t}{2}\right)\right]^{-1},
\nonumber
\\
a&=& a_0 \left(H_0+ \frac{3}{\Gamma_0} \exp \left(\frac{\gamma \Gamma_0 t}{2}\right)\right)^{\frac{2}{3 \gamma}},
\nonumber
\\
\rho (t)&=& 3 \exp \left(\gamma \Gamma_0 t \right) \left[H_0+ \frac{3}{\Gamma_0} \exp \left(\frac{\gamma \Gamma_0 t}{2}\right)\right]^2,
\nonumber
\\
\rho_{in} (t, r)&=& -\left(\frac{r}{r_0}\right)^{-\frac{1+3 \omega_r}{\omega_r}} \left[r_0 ^2 a_0 ^2 \omega_r \left(H_0+ \frac{3}{\Gamma_0} \exp \left(\frac{\gamma \Gamma_0 t}{2}\right)\right)\right]^{-1}.
\end{eqnarray}

These solutions represent an evolving wormhole having throat at $r_0$ provided $\omega_r< -1$ or $\omega_r> 0$ \cite{Cataldo2} and asymptotically it describes a flat FRW universe. It should be noted that in general to keep a wormhole open, an exotic matter with negative energy density is needed \cite{Morris1, Visser1}, although it is possible to have evolving wormholes satisfying the dominant energy condition (DEC), so that the energy density is positive everywhere \cite{Maeda1, Cataldo2}. In the above solution, if $\omega_r> 0$, then the inhomogeneous matter component (threading the wormhole) has positive radial pressure but the energy density $\rho_{in}$ and transverse pressure are negative while for $\omega_r< -1$, the situation is reversed, i.e., $\rho_{in}, \rho_t> 0$ and $p_r< 0$ with $|p_r|> \rho_{in}$. However, the total energy density defined by

$$\rho_{tot}= \rho (t)+ \rho_{in} (t, r),$$

is positive definite for $\omega_r< 0$ throughout the evolution, but, for $\omega_r> 0$, positivity of $\rho_T$ is confined to some time interval. In particular, if $\gamma< 2/3$, $\rho_{tot}> 0$ for $t> t_0$, while if $\gamma> 2/3$, $\rho_{tot}< 0$ for $t> t_0$; where $t_0$ is given by the following equation

\begin{equation}
3 \tilde{T}^2 \left(H_0+ \frac{3 \tilde{T}}{\Gamma_0}\right)^{\frac{4}{3 \gamma}- 2}= \frac{1}{r_0 ^2 a_0 ^2 \omega_r} \left(\frac{r}{r_0}\right)^{-\frac{1+ \omega_r}{\omega_r}},~~~with~~\tilde{T}= e^{\frac{\gamma \Gamma_0 t}{2}}.
\end{equation}

For $\gamma= \frac{2}{3}$, $\rho_{tot}> 0$ for $t> t_1$; where $t_1$ has the expression

\begin{equation}
t_1= \frac{3}{2 \Gamma_0} ln \left[\frac{(\frac{r}{r_0})^{-\frac{1+ \omega_r}{\omega_r}}}{3 r_0 ^2 a_0 ^2 \omega_r}\right].
\end{equation}

In particular, if the expansion occurs at a constant velocity for $\gamma= 2/3$, then both the matter components evolves as $1/a^2$. It should be mentioned that $\gamma< 2/3$ or $\gamma> 2/3$ corresponds to accelerating or decelerating phase of the evolution.\\

Further, at $t= 0$, the total energy density defined by $\rho_ {{tot}_{0}} (r)$ is given by

\begin{equation}
\rho_ {{tot}_{0}} (r)= \frac{3}{(H_0+ \frac{3}{\Gamma_0})^2}- \frac{1}{r_0 ^2 a_0 ^2 \omega_r (H_0+ \frac{3}{\Gamma_0})^{\frac{4}{3 \gamma}}} \left(\frac{r}{r_0}\right)^{-\frac{1+ 3 \omega_r}{\omega_r}}.
\end{equation}

Hence the homogeneous and isotropic fluid density exceeds (in magnitude) the energy density of the other fluid component for the following restrictions:

$$r> \left[\frac{\left(H_0+ \frac{3}{\Gamma_0}\right)^{2- \frac{4}{3 \gamma}}}{3 r_0 ^{-1-\frac{1}{\omega_r}} a_0 ^2 \omega_r}\right]^{\frac{\omega_r}{1+ 3 \omega_r}},~~~~~~~~when~\omega_r > 0,$$

$$r< \left[\frac{3 r_0^{-1- \frac{1}{\omega_r}} a_0^2 |\omega_r|}{\left(H_0+ \frac{3}{\Gamma_0}\right)^{2- \frac{4}{3 \gamma}}}\right]^{|\frac{\omega_r}{1+ 3 \omega_r}|},~~~~~~when~-\frac{1}{3}< \omega_r< 0,$$

$$r> \left[\frac{\left(H_0+ \frac{3}{\Gamma_0}\right)^{2- \frac{4}{3 \gamma}}}{3 r_0^{-1- \frac{1}{\omega_r}} a_0^2 |\omega_r|}\right]^{\frac{\omega_r}{1+ 3 \omega_r}} when~\omega_r< -\frac{1}{3}.$$

We now define the notion of equilibrium time ($t_{eq}$) as the instant when both the matter components have equal energy density, i.e., $\rho (t_{eq})= \rho_{in} (t_{eq}, r)$.\\

From Eq. (36) we have $\tilde{T}_{eq}(= e^{\gamma \Gamma_0 \frac{t_{eq}}{2}})$ as the positive root of the equation (in $\tilde{T}$)

\begin{equation}
3 \tilde{T}^2 \left(H_0+ \frac{3 \tilde{T}}{\Gamma_0}\right)^{\frac{4}{3 \gamma}- 2}= -\frac{1}{r_0 ^2 a_0 ^2 \omega_r} \left(\frac{r}{r_0}\right)^{-\frac{1+ \omega_r}{\omega_r}}.
\end{equation}

In particular, for $\gamma= \frac{2}{3}$, $t_{eq}$ exists only for $\omega_r< 0$, and is given by

\begin{equation}
t_{eq}= \frac{3}{2 \Gamma_0} ln \left[\frac{\left(\frac{r}{r_0}\right)^{-\frac{1+ \omega_r}{\omega_r}}}{3 r_0 ^2 a_0 ^2 \omega_r}\right].
\end{equation}

Lastly, note that, if $H_0< 0$ and $\gamma< 0$ (i.e., the homogeneous fluid is in the phantom domain), then at finite time,

$$t_{s_1}= \frac{2}{|\gamma| \Gamma_0} ln |\frac{3}{H_0 \Gamma_0}|,~~a\longrightarrow \infty;~ \rho(t)\longrightarrow \infty;~ p\longrightarrow -\infty;~ \rho_{in} \longrightarrow 0;~ \Pi \longrightarrow - \infty.$$

So, we have a future singularity (big rip) at a finite value of the co-moving proper time $t_{s_1}$. At the time of singularity, the anisotropic matter threading the wormhole vanishes while there is constant particle creation rate throughout the evolution. However, for $\gamma= 0$, the scale factor has the exponential form as

\begin{equation}
a= a_0 \exp\left[\frac{2 H_0}{3 \Gamma_0} \exp \left(\frac{3 \Gamma_0 t}{2}\right)\right].
\end{equation}

which clearly shows that the evolution does not end in a future singularity rather there will be an accelerated expansion.

$~~~~~~~~~~~~~~~~~~~~~~~~~~~~~~~~~~~~~~~~~~~~~~~~~~${{\bf Case II:} $\Gamma= \Gamma_0 H$}

Another accelerating wormhole solution is obtained when the particle creation rate is proportional to the Hubble parameter. Solving the evolution Eq. (33) we obtain

\begin{eqnarray}
H&=& \frac{H_0}{\left(1+ \frac{H_0 \gamma}{2} (3- \Gamma_0) t\right)},
\nonumber
\\
a&=& a_0 \left(1+ \frac{H_0 \gamma}{2} (3- \Gamma_0) t\right)^{\frac{2}{\gamma (3- \Gamma_0)}},
\nonumber
\\
\rho (t)&=& \frac{3 H_0 ^2}{\left(1+ \frac{H_0 \gamma}{2} (3- \Gamma_0) t\right)^2},
\nonumber
\\
\rho_{in} (t, r)&=& -\frac{\left(\frac{r}{r_0}\right)^{-(1+ \omega_r)/\omega_r}}{r_0 ^2 \omega_r a_0 ^2 \left[\left(1+ \frac{H_0 \gamma}{2} (3- \Gamma_0) t\right)^{\frac{4}{\gamma (3- \Gamma_0)}}\right]},
\nonumber
\\
\Gamma&=& \frac{\Gamma_0 H_0}{\left(1+ \frac{H_0 \gamma}{2} (3- \Gamma_0) t \right)}.
\end{eqnarray}

From the above solution one may notice that if $\gamma (3- \Gamma_0)< 0$ and $H_0> 0$, the scale factor, isotropic energy density and the pressure diverges at a finite time

\begin{equation}
t_{s_2}= \frac{2}{H_0 |\gamma (3- \Gamma_0)|}.
\end{equation}

Also, at this time instant, the inhomogeneous matter density and anisotropic pressure threading the wormhole vanishes for $r\geq r_0$; so here the dissipative expanding wormhole is also associated with a future singularity which is also big rip in nature. Interestingly, one may note that in the above scenario we have the restriction: $\gamma (3- \Gamma_0)< 0$ which implies either $\gamma< 0$, $\Gamma_0< 3$ or $\gamma> 0$, $\Gamma_0> 3$. Hence big rip singularity occurs not only for phantom dissipative fluid but also it is possible for dissipative dark energy or even for normal dissipative fluid. Here, if we consider the models expanding with constant velocity, then from (32) (with $\Lambda= 0$) $\gamma$ is restricted by the following relation

\begin{equation}
\frac{\Gamma_0 \gamma}{3}= 3 \gamma- 2, ~~~i.e.,~~6= \gamma (9- \Gamma_0).
\end{equation}

Due to this restriction, the above two possibilities can be restated as---(i) $\gamma< 0$, $\Gamma_0< 0$, or (ii) $\gamma> 0$, $3< \Gamma_0< 9$. Hence for the expansion with constant velocity either there is particle annihilation for the dissipative phantom isotropic fluid or we have dissipative dark energy or normal fluid with particle creation mechanism. The total matter density for the wormhole solution (43) is given by

\begin{equation}
\rho_{tot} (t, r)= \frac{3 H_0 ^2}{\left(1+ \frac{H_0 \gamma (3-\Gamma_0) t}{2}\right)^2}- \frac{(\frac{r}{r_0})^{-\frac{1+3 \omega_r}{\omega_r}}}{r_0 ^2 a_0 ^2 \omega_r} \left(1+ \frac{H_0 \gamma (3-\Gamma_0) t}{2}\right)^{-\frac{4}{\gamma (3- \Gamma_0)}},
\end{equation}

which is clearly positive definite for all $r$, if $\omega_r< 0$. But, for $\omega_r> 0$, $\rho (t)$ dominates initially till $t_0$ (say), then the inhomogeneous fluid component takes the leading role. Initially, at $t= 0$ the total energy density has the expression

\begin{equation}
\rho_{{tot}_{0}}= 3 H_0^2- \frac{1}{r_0^2 a_0^2 \omega_r} \left(\frac{r}{r_0}\right)^{-\frac{1+3 \omega_r}{\omega_r}}.
\end{equation}

So, initial isotropic matter density will dominate (in magnitude) over the inhomogeneous component, provided, the radial co--ordinate $r$ has the following restrictions:

$$r> \left(\frac{\omega_r^{-1} a_0^{-2}}{3 H_0^2} r_0^{1+ \frac{1}{\omega_r}}\right),~~~~~~~if~\omega_r> 0,$$

$$r< \left(3 H_0^2 |\omega_r| a_0^2 r_0^{-(1+\frac{1}{\omega_r})}\right),~~~~~~if~-\frac{1}{3}< \omega_r< 0,$$

$$r> \left(\frac{1}{3 H_0 ^2 |\omega_r| a_0^2} r_0^{1+ \frac{1}{\omega_r}}\right),~if~\omega_r< - \frac{1}{3},$$

Also, for the present model, the equilibrium time configuration can be obtained as

\begin{equation}
t_{eq}= \frac{2}{\gamma H_0 (3-\Gamma_0)} \left[\left(-\frac{3 H_0 ^2 \omega_r a_0 ^2 r_0 ^2}{(\frac{r}{r_0})^{-\frac{1+ 3 \omega_r)}{\omega_r}}}\right)^{\frac{\gamma (3-\Gamma_0)}{2 \left(2+ \gamma (3-\Gamma_0)\right)}} -1 \right],
\end{equation}

which can take complex values. However, it is always real, provided, $\omega_r< 0$ and for positive definiteness we have the restriction:

\begin{equation}
\left[\frac{3 H_0 ^2 |\omega_r| a_0 ^2 r_0 ^2}{(\frac{r}{r_0})^{-\frac{1+ 3 \omega_r)}{\omega_r}}}\right]^{\frac{\gamma (3-\Gamma_0)}{2 \left(2+ \gamma (3-\Gamma_0)\right)}} < 1.
\end{equation}

Moreover, it looks interesting to note that if the isotropic fluid satisfies DEC, i.e., $0< \gamma< 2$, then one can rescale the cosmic time so that $1+ \frac{H_0 \Gamma}{2} (3- \Gamma_0) t \longrightarrow t$, provided $H_0 (3- \Gamma_0)> 0$. As a result, the scale factor has the usual power law form $a(t)= a_0 t^{\frac{2}{\gamma (3- \Gamma_0)}}$ with isotropic energy density $\rho(t)= 3 H_0 ^2/t^2$, and hence there is no future singularity.\\

$~~~~~~~~~~~~~~~~~~~~~~~~~~~~~~~~~~~~~~~~~~~~~~~~~~${{\bf Case III:} $\Gamma= 3\frac{H_0}{H}$}

This choice of the particle creation rate results the following cosmological solutions corresponding to an expanding wormhole configuration:

$$a= \frac{a_0}{(H_0)^\frac{1}{3 \gamma}} \left[\sinh \left(\frac{3 \gamma}{2} \sqrt{H_0} (t- t_0)\right)\right]^{\frac{2}{3 \gamma}},$$

$$\rho (t)= 3 H_0 \coth^2 \left[\frac{3 \gamma}{2} \sqrt{H_0} (t- t_0)\right],$$

$$\Gamma= 3 \sqrt{H_0} \tanh \left[\frac{3 \gamma}{2} \sqrt{H_0} (t- t_0)\right],$$

\begin{equation}
\rho_{in}= -\left(\frac{r}{r_0}\right)^{-\frac{1+ 3 \omega_r}{\omega_r}} (H_0)^\frac{2}{3 \gamma} \frac{1}{r_0 ^2 a_0^2 \omega_r} \left[\sinh \left(\frac{3 \gamma}{2} \sqrt{H_0} (t- t_0)\right)\right]^{-\frac{4}{3 \gamma}}.
\end{equation}

Here the integration constant $t_0$ ($> 0$) corresponds to a future singularity provided $\gamma< 0$, i.e., the homogeneous fluid is phantom in nature. At the singularity, the scale factor, isotropic matter density, thermodynamic and dissipative pressure all blow up to infinity, only the inhomogeneous matter density, anisotropic pressure and the particle cfreation rate vanish. So as in the previous cases this wormhole solution also corresponds to a future big rip singularity. As in the previous two cases, if $\omega_r< 0$, then the total energy density is positive throughout the evolution, but for $\omega_r> 0$, the isotropic energy density dominates over the inhomogeneous matter component (for a constant $r$) provided,

\begin{equation}
\frac{\cosh^2 \left[\frac{3 \gamma}{2} \sqrt{H_0} (t-t_0)\right]}{\left[\sinh(\frac{3 \gamma}{2} \sqrt{H_0} (t-t_0))\right]^{2-\frac{4}{3 \gamma}}}> -\frac{(\frac{r}{r_0})^{-\frac{1+ 3 \omega_r}{\omega_r}}}{3 H_0 r_0 ^2 a_0 ^2 \omega_r}.
\end{equation}

Otherwise, the inhomogeneous energy density has the dominating role. Lastly, the notion of equilibrium time ($t_{eq}$) will be realistic provided $\omega_r< 0$, and it is characterized by the following equation

\begin{equation}
\frac{\cosh^2 \left(\frac{3 \gamma}{2} \sqrt{H_0} (t- t_0)\right)}{\left[\sinh \left(\frac{3 \gamma}{2} \sqrt{H_0} (t- t_0)\right)\right]^{2- \frac{4}{3 \gamma}}}= \frac{r_0^{1+\frac{1}{\omega_r}}}{3 H_0 a_0^2 |\omega_r|} r^{-\frac{1+3 \omega_r}{\omega_r}}.
\end{equation}

\section{Discussions and Final Remarks}

In the present work we deal with a FRW like space--time model which is both inhomogeneous and anisotropic in nature and there is a future singularity at a finite proper time. There are two non-interacting matter components: One is isotropic and homogeneously distributed dissipative fluid, while the other matter component is both inhomogeneous and anisotropic in nature. Here, we have considered dissipation due to particle creation mechanism and for simplicity we restrict ourselves to adiabatic process so that the dissipative pressure is linearly related to the particle creation rate. The solutions presented in the paper describe evolving wormholes which are threaded and sustained by the inhomogeneous and anisotropic fluid component, while the rate of expansion is characterized by the isotropic and homogeneous matter part which in most of the cases is chosen in the phantom domain. Three cosmological models are presented in the paper corresponding to three different choices of the particle creation rates, and in all of them, the wormhole models encounter a big rip singularity in course of its evolution. As all the wormhole solutions are asymptotically flat FRW cosmologies, so all the results on future singularities obtained are also true for flat FRW cosmological models.\\

Here, the dissipation is chosen as bulk viscosity and the corresponding pressure is related to the particle creation mechanism. As in the literature \cite{Barrow1}, the bulk viscosity is in the power law form of the Hubble parameter, so, due to the isentropic condition (Eq. (28)), it is reasonable to choose the particle creation rate as some power of the Hubble parameter. For simplicity, we have restricted to: (i) $\Gamma=$ constant \cite{Chakraborty2}, (ii) $\Gamma \propto H$ \cite{Zimdahl1}, and (iii) $\Gamma \propto 1/H$ \cite{Chakraborty3}. The solutions corresponding to the first and third choices, the homogeneous and isotropic fluid is always phantom in nature, but for the second choice, there are two possibilities of which one is similar as the other choices while for the second possibility, there exists evolving wormhole without phantom energy. Hence one can say that phantom energy is not essential for describing evolving wormholes.\\

Further, it should be noted that in the present work the matter component is chosen such that there is no mixed component of the energy-momentum tensor (i.e., $T_{rt}= 0$). As a result, neither there is any radial energy flow, nor there is any accretion onto the wormhole from the cosmic fluid. So, we may conclude that the present model can not be able to explain the big trip mechanism.

\section{Acknowledgements}

SP thanks CSIR, Govt. of India for research grants through SRF scheme (File No: 09/096 (0749)/2012-EMR-I). SC thanks UGC-DRS programme, Department of Mathematics, Jadavpur University. Also, a warm thanks goes to Inter University Centre for Astronomy and Astrophysics (IUCAA), Pune, India on behalf of SC for their hospitality as the initiation of this work was taken during a visit there.

\end{document}